\documentstyle[twoside,fleqn,espcrc2,epsfig]{article}

\newcommand{\WIDTH}{7.5cm}
\title{\makebox[0pt][l]{\raisebox{3cm}[0pt][0pt]{%
\parbox{5cm}{\normalsize DESY 97--258 \\ FUB-HEP/97--07 \\
HUB-HEP--97/56 \\ September 1997}}}%
Resolution of the Landau pole problem in QED\thanks{Talk
presented by H. St\"uben}}

\author{%
M. G\"ockeler\address{Institut f\"ur Theoretische Physik,
Universit\"at Regensburg, D-93040 Regensburg},
R. Horsley\address{Institut f\"ur Physik, Humboldt Universit\"at zu
Berlin, D-10115 Berlin},
V. Linke\address{Institut f\"ur Theoretische Physik, Freie
Universit\"at Berlin, D-14195 Berlin},
P. Rakow\address{Deutsches Elektronen Synchrotron DESY, Institut f\"ur
Hochenergiephysik und HLRZ, D-15735 Zeuthen},
G. Schierholz$^{\rm d,}$\address{Deutsches Elektronen Synchrotron DESY,
D-22603 Hamburg}
and H. St\"uben\address{Konrad-Zuse-Zentrum f\"ur Informationstechnik
Berlin, D-14195 Berlin}
}

\begin{document}

\begin{abstract}
We present new numerical results for the renormalized mass and
coupling in non-compact lattice QED with staggered fermions.
Implications for the continuum limit and the role of the 
Landau pole are discussed.
\end{abstract}

\maketitle

\section{INTRODUCTION}

In the 1950s Landau investigated the relation between the bare charge
$e$ and the renormalized charge $e_R$ in QED.  He found \cite{Landau}
\begin{equation}
   \frac{1}{e_R^2} - \frac{1}{e^2} = 
   \frac{N_f}{6 \pi^2} \, \ln \frac{\Lambda}{m_R}\,,
   \label{eq-landau}
\end{equation}
where $\Lambda$ is the momentum cutoff, $m_R$ is the renormalized mass
of the electron and $N_f$ is the number of flavours (for staggered
fermions $N_f = 4$).  It is well known that (\ref{eq-landau}) implies
two potential problems when $\Lambda \rightarrow \infty$:
\begin{itemize}
\item When $e$ is fixed the theory has a {\em trivial} continuum
limit, i.e., $e_R \rightarrow 0$.

\item When $e_R > 0$ is fixed $e$ becomes singular at 
$\Lambda_{\rm Landau} = m_R \exp{(6 \pi^2/N_f e_R^2)}$.
This singularity is called the {\em Landau pole}.
\end{itemize}
These problems do not affect the phenomenological success of QED
because in practical perturbative calculations the cutoff can be
chosen to be large compared with experimental energies.  Finding a
solution to them is of fundamental theoretical interest and requires
non-perturbative methods.  We have therefore extensively studied QED
on the lattice using non-compact gauge-fields and dynamical staggered
fermions \cite{QED,EOS}.

In this talk we report on new measurements of the renormalized mass
and charge.  This was started in \cite{QED} and is now extended to
lattices of size $16^4$ and bare masses $am$ down to 0.005 ($a$
denotes the lattice spacing).  Using our measurements we have
determined {\em functions\/} $am_R(e^2, am)$ and $e_R^2(e^2, am)$.
These functions imply that the theory is trivial. They also give a
resolution of the Landau pole problem.

\begin{figure*}[t]
\begin{picture}(16,7)(0,0.5)
\put(0.0,0){\epsfig{file=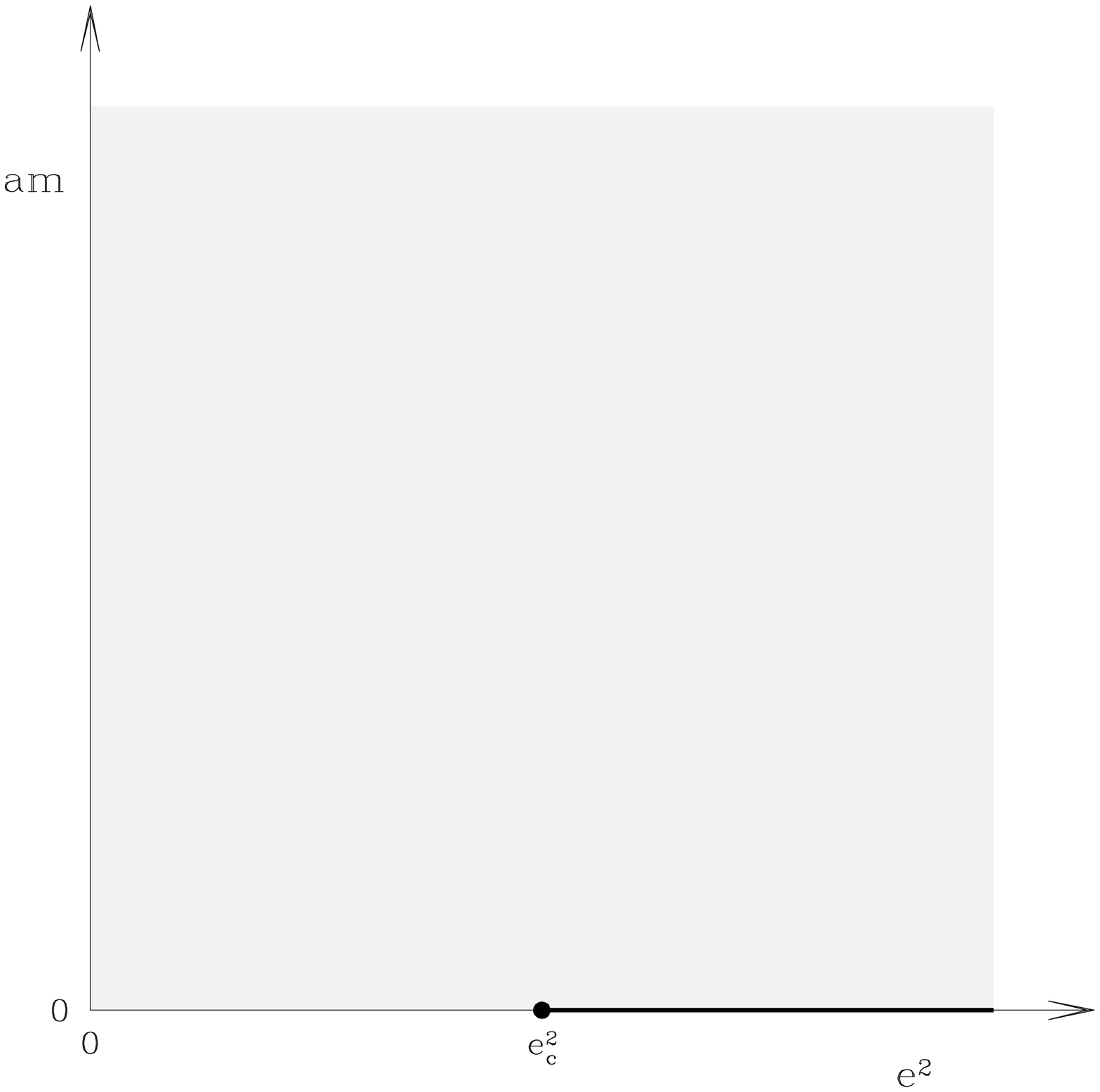,width=\WIDTH}}
\put(8.5,0){\epsfig{file=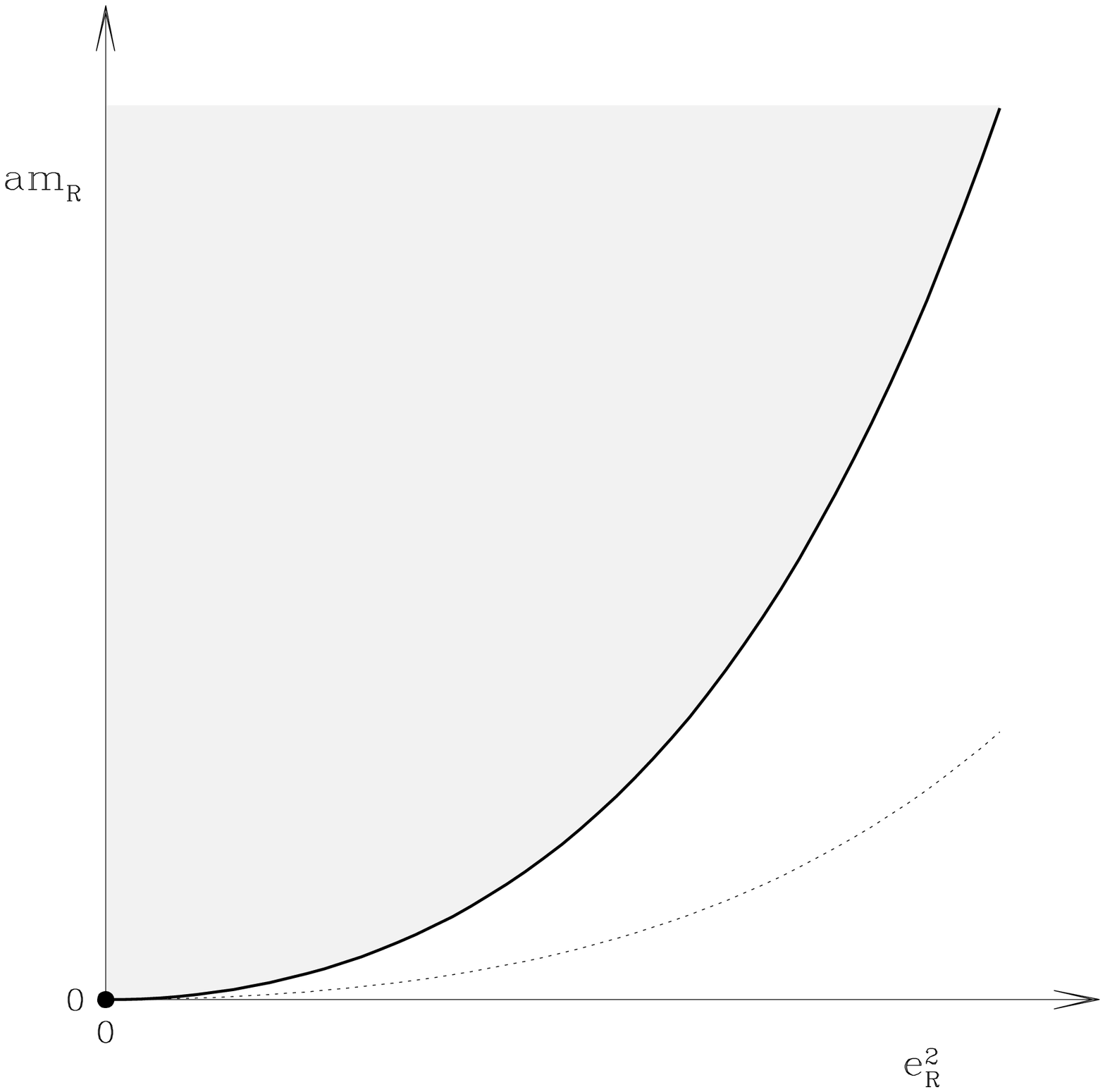,width=\WIDTH}}
\end{picture}
\caption{Sketch of the mapping $(e^2, am) \leftrightarrow (e_R^2, am_R)$.}
\label{fig-e-mass}
\end{figure*}

\section{THE RENORMALIZED MASS}

The renormalized mass was obtained from fits to the fermion propagator
as explained in \cite{QED}.  To get $am_R$ as a function of $am$ and
$e$ we need to know the equation of state which relates the bare
parameters to the chiral condensate $\sigma \equiv a^3 \langle
\bar{\chi}\chi \rangle$.  In \cite{EOS} we have found that the
$\sigma$ data obey a mean field equation of state with logarithmic
corrections
\begin{equation}
 am = A_0 \frac{\sigma^3}{\ln^{p_0}(1/\sigma)}
   + A_1 \!\left(\frac{1}{e^2} - \frac{1}{e_c^2}\right) 
     \frac{\sigma}{\ln^{p_1}(1/\sigma)}\,.
\end{equation}
We fitted this expression to $\sigma$ data that were
extrapolated to infinite lattice size and obtained $1/e_c^2 =
0.19040(9)$, $A_0 = 1.798(5)$, $p_0 = 0.324(15)$, $A_1 = 6.76(3)$,
$p_1 = 0.485(7)$, $\chi^2/{\rm d.o.f} = 7.6$.

We have observed that $\sigma$ can be well described by a polynomial in
$am_R$
\begin{equation}
  \sigma = A_1 a m_R + A_3 a^3 m_R^3 + A_5 a^5 m_R^5 + A_7 a^7 m_R^7
  \label{eq-poly}
\end{equation}
where the first parameter $A_1 \equiv 0.6197$ can be taken from
perturbation theory and a fit to data from $12^4$ and $16^4$ lattices
gave $A_3 = -0.321(5)$, $A_5 = +0.169(13)$, $A_7 = -0.040(7)$,
$\chi^2/{\rm d.o.f.} = 2.1$. Because the results of both lattice sizes
fall on a universal curve we conclude that the polynomial
(\ref{eq-poly}) is also valid on an infinite lattice.

\section{THE RENORMALIZED CHARGE}

The determination of the renormalized charge has been
improved since \cite{QED}.  The method can only be sketched
here.  It consists of making a global fit to {\em all\/} gauge field
propagators $D(k)$ that we have measured
\begin{equation}
   \frac{1}{e^2 D(k)} - \frac{1}{e^2} = -\Pi(k, am_R, L)
   \label{eq-gfit}
\end{equation}
where $L$ is the linear lattice size.  The {\em ansatz\/} for the
fit function $\Pi$ was taken from \cite{Shirkov} to be
\begin{equation}
 \Pi =  U - \frac{V}{U} \ln(1 - e^2 U)
\end{equation}
where $U$ is given by 1-loop lattice perturbation theory (see
\cite{QED}) and where we have set
\begin{equation}
  V(k, am_R, L) = v_0 + v_1 U(k, am_R, L) \,.
\end{equation}
We then find $e_R^2(e^2, am_R)$ using $e_R^2 = Z_3 e^2$ and $Z_3 =
\lim_{k \rightarrow 0} \lim_{L \rightarrow \infty} D(k)$ from
\begin{equation}
   \frac{1}{e_R^2} - \frac{1}{e^2} = - \Pi(0, am_R, \infty)\,.
   \label{eq-solution}
\end{equation}
A simultaneous fit of (\ref{eq-gfit}) to the gauge field propagators
at our 52 values $(e^2, am, L)$ gave $v_0 = -0.00207(2)$, $v_1 =
-0.0328(7)$, $\chi^2/{\rm d.o.f.} = 1.7$.  Since $U(0, am_R, \infty)
\approx (N_f/6\pi^2) (-0.31 + \ln am_R)$ we only find small
corrections to the old result (\ref{eq-landau}).

\begin{figure*}[t]
\begin{picture}(16,7)(0,0.5)
\put(0.0,0){\epsfig{file=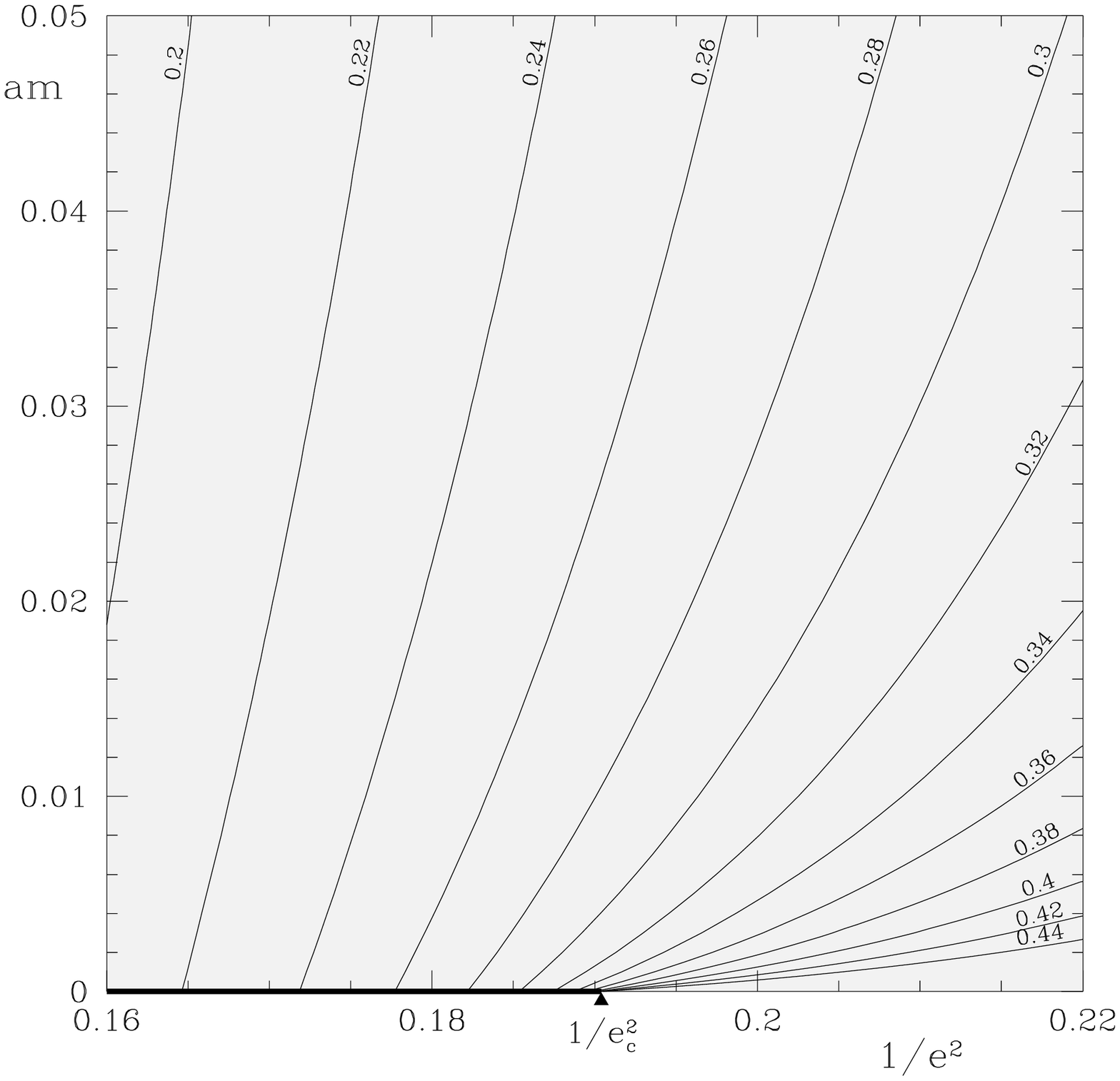,width=\WIDTH}}
\put(8.5,0){\epsfig{file=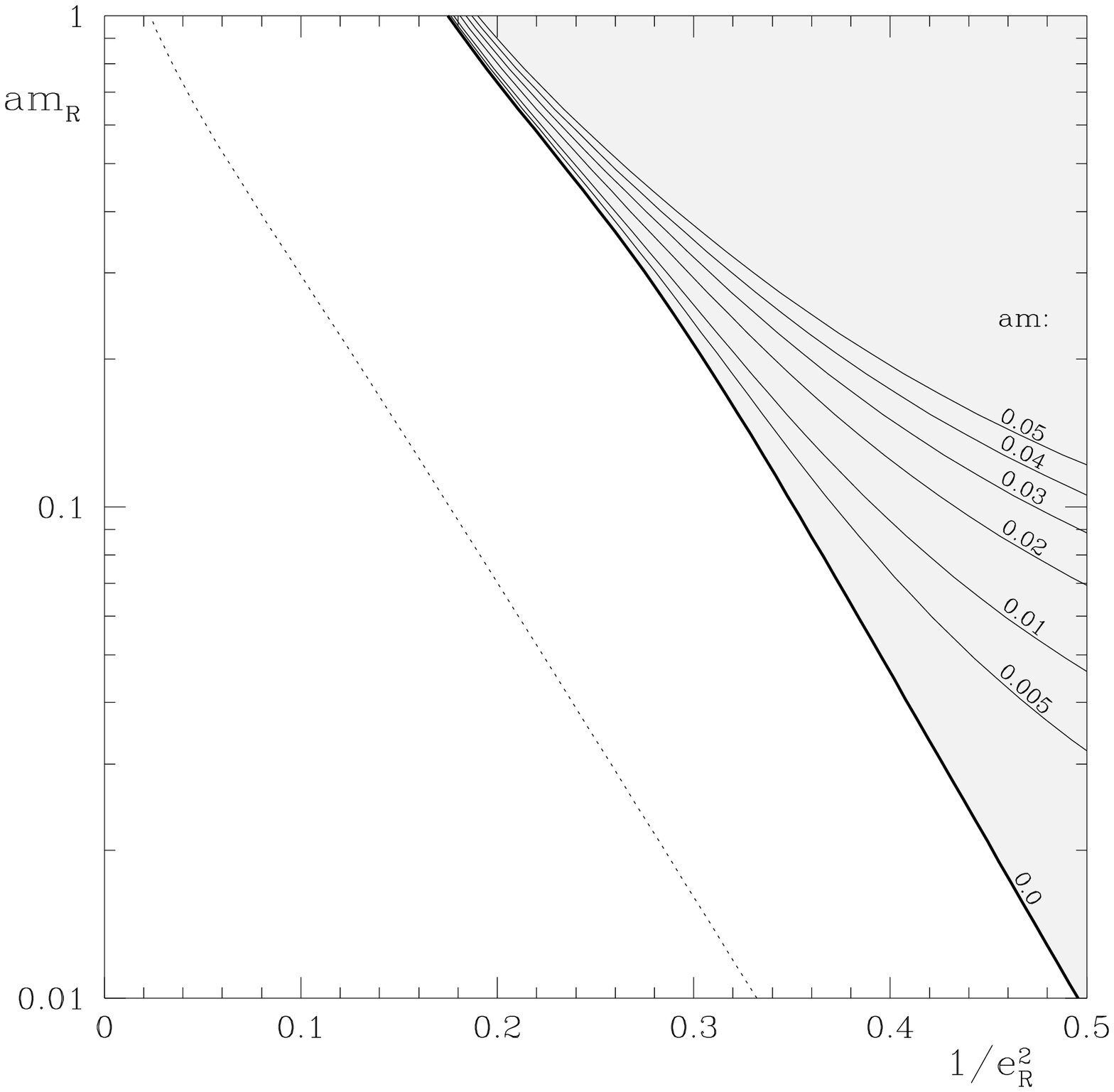,width=\WIDTH}}
\end{picture}
\caption{Quantitative picture of the mapping $(1/e^2, am)
\leftrightarrow (1/e_R^2, am_R)$. In the plane of the bare parameters
lines of constant $1/e_R^2$ are shown indicating the
renormalization group flow of $e_R^2$. In the plane of the
renormalized parameters the lines of constant $am$ are shown.}
\label{fig-b-mass}
\end{figure*}

\section{DISCUSSION}

We can now discuss the mapping $(e^2, am) \leftrightarrow (e_R^2,
am_R)$. A global qualitative view of this mapping is shown in Figure
\ref{fig-e-mass}, while a quantitative plot of $(1/e^2, am)
\leftrightarrow (1/e_R^2, am_R)$ is shown in Figure \ref{fig-b-mass}.

In both figures accessible regions are plotted in grey. The whole plane
of bare parameters is accessible but this plane is mapped only onto a
part of the plane of the renormalized parameters. The border of the
accessible region is shown as a thick line. It is the image of the
corresponding thick line on the $am = 0$ line starting/ending at the critical
value of the coupling constant. 

On the line $am_R = 0$ in Figure \ref{fig-e-mass} the only accessible point
is the origin. This reflects triviality. In Figure
\ref{fig-b-mass} triviality is expressed by the fact that no line of
(finite) constant $1/e_R^2$ flows into the critical point.

The dotted line in both figures is the position of the Landau pole,
i.e., the line of pairs $(e_R^2, am_R)$ with $1/e^2 = 0$ from
(\ref{eq-solution}) with $\Pi \equiv U$. This line is well separated
from the border for all finite $e_R^2$.

\section{CONCLUSIONS}

{}From the presented analysis we conclude that non-compact lattice QED
with staggered fermions has a trivial continuum limit. In addition our
analysis implies a resolution of the Landau problem. The resolution
is that for given $e$ and $am$ the theory does not allow arbitrary
choices of $am_R$.  Instead through chiral symmetry breaking the theory
itself provides a minimal lattice spacing or maximal cutoff which is
below $\Lambda_{\rm Landau}$. Non-perturbatively the function $\Pi$ is
very close to what one finds in perturbation theory, but there is no
Landau pole problem because the pole always lies in the forbidden
region.

\section*{ACKNOWLEDGEMENTS}

Our new measurements were done on the CRAY T3D at the
Konrad-Zuse-Zentrum Berlin.  Financial support by the Deutsche
Forschungs\-gemeinschaft is gratefully acknowledged.

\end{document}